\documentclass[article,twocolumn,showpacs,floatfix,
footinbib,superscriptaddress]{revtex4}

\usepackage{graphicx}
\usepackage{epstopdf}
\usepackage{color}

\begin{document}

\title{CAST solar axion search with $^3$He buffer gas: Closing the hot dark matter gap}

\newcommand{\Dogus}{Dogus University, Istanbul, Turkey}
\newcommand{\Saclay}{IRFU, Centre d'\'Etudes Nucl\'eaires de Saclay (CEA-Saclay), Gif-sur-Yvette, France}
\newcommand{\CERN}{European Organization for Nuclear Research (CERN), Gen\`eve, Switzerland}
\newcommand{\INR}{Institute for Nuclear Research (INR), Russian Academy of Sciences, Moscow, Russia}

\newcommand{\MPE}{Max-Planck-Institut f\"{u}r Extraterrestrische Physik, Garching, Germany}
\newcommand{\Trieste}{Istituto Nazionale di Fisica Nucleare (INFN), Sezione di Trieste and Universit\`a di Trieste, Trieste, Italy}
\newcommand{\Zaragoza}{Grupo de Investigaci\'{o}n de F\'{\i}sica Nuclear y Astropart\'{\i}culas, Universidad de Zaragoza, Zaragoza, Spain }
\newcommand{\Chicago}{Enrico Fermi Institute and KICP, University of Chicago, Chicago, IL 60637, USA}
\newcommand{\Thessaloniki}{Aristotle University of Thessaloniki, Thessaloniki, Greece}
\newcommand{\Demokritos}{National Center for Scientific Research ``Demokritos'', Athens, Greece}
\newcommand{\Freiburg}{Albert-Ludwigs-Universit\"{a}t Freiburg, Freiburg, Germany}
\newcommand{\Patras}{Physics Department, University of Patras, Patras, Greece}
\newcommand{\Athens}{National Technical University of Athens, Athens, Greece}
\newcommand{\MPI}{MPI Halbleiterlabor, M\"{u}nchen, Germany}
\newcommand{\Vancouver}{Department of Physics and Astronomy, University of British Columbia, Vancouver, Canada }
\newcommand{\Darmstadt}{Technische Universit\"{a}t Darmstadt, IKP, Darmstadt, Germany}
\newcommand{\Frankfurt}{Johann Wolfgang Goethe-Universit\"at, Institut f\"ur Angewandte Physik, Frankfurt am Main, Germany}
\newcommand{\Zagreb}{Rudjer Bo\v{s}kovi\'{c} Institute, Zagreb, Croatia}
\newcommand{\MPP}{Max-Planck-Institut f\"{u}r Physik (Werner-Heisenberg-Institut), M\"unchen, Germany}
\newcommand{\LLNL}{Lawrence Livermore National Laboratory, Livermore, CA 94550, USA}
\newcommand{\MPIS}{Max-Planck-Institut f\"{u}r Sonnensystemforschung, G\"{o}ttingen, Germany}
\newcommand{\Bogazici}{Bogazici University, Istanbul, Turkey.}
\newcommand{\Manchester}{School of Physics and Astronomy, Schuster Laboratory, University of Manchester, Manchester, UK.}
\newcommand{\PNSensor}{PNSensor GmbH, M\"unchen, Germany.}
\newcommand{\XFEL}{European XFEL GmbH, Notkestrasse 85, 22607 Hamburg, Germany.}
\newcommand{\ECU}{Excellence Cluster Universe, Technische Universit\"{a}t M\"unchen, Garching, Germany.}
\newcommand{\Berkeley}{University of California Berkeley, CA 94720, USA.}
\newcommand{\BenGurion}{Physics Department, Ben-Gurion University of the Negev, Beer-Sheva 84105, Israel.}
\newcommand{\Korea}{School of Space Research, Kyung Hee University, Yongin, Republic of Korea.}
\newcommand{\ESS}{European Spallation Source ESS AB, Lund, Sweden.}
\newcommand{\LAL}{Laboratoire de l'Acc\'{e}l\'{e}rateur Lin\'{e}aire (LAL), Orsay, France.}
\newcommand{\Rijeka}{Physics Department and Center for Micro and Nano Sciences and Technologies, University of Rijeka, Radmile Matejcic 2, 51000 Rijeka, Croatia.}
\newcommand{\EPFL}{Laboratoire de Transfert de Chaleur et de Masse, \'Ecole Polytechnique F\'ed\'erale de Lausanne (EPFL), Lausanne, Switzerland.}

\author{    M.~Arik}\altaffiliation[Present addr.: ]{\Bogazici}\affiliation{\Dogus}
\author{    S.~Aune  }\affiliation{\Saclay}
\author{    K.~Barth  }\affiliation{\CERN}
\author{    A.~Belov  }\affiliation{\INR}
\author{    S.~Borghi  }\altaffiliation[Present addr.: ]{\Manchester}\affiliation{\CERN}
\author{    H.~Br\"auninger  }\affiliation{\MPE}
\author{    G.~Cantatore  }\affiliation{\Trieste}
\author{    J.~M.~Carmona  }\affiliation{\Zaragoza}
\author{    S.~A.~Cetin  }\affiliation{\Dogus}
\author{    J.~I.~Collar  }\affiliation{\Chicago}
\author{    E.~Da~Riva  }\affiliation{\CERN}
\author{    T.~Dafni  }\email[Corresponding author: ]{Theopisti.Dafni@unizar.es}\affiliation{\Zaragoza}
\author{    M.~Davenport  }\affiliation{\CERN}
\author{    C.~Eleftheriadis  }\affiliation{\Thessaloniki}
\author{    N.~Elias  }\altaffiliation[Present addr.: ]{\ESS}\affiliation{\CERN}
\author{    G.~Fanourakis  }\affiliation{\Demokritos}
\author{    E.~Ferrer-Ribas  }\affiliation{\Saclay}
\author{    P.~Friedrich    }\affiliation{\MPE}
\author{    J.~Gal\' an  }\affiliation{\Zaragoza}\affiliation{\Saclay}
\author{    J.~A.~Garc\' ia  }\affiliation{\Zaragoza}
\author{    A.~Gardikiotis  }\affiliation{\Patras}
\author{    J.~G.~Garza  }\affiliation{\Zaragoza}
\author{    E.~N.~Gazis  }\affiliation{\Athens}
\author{    T.~Geralis  }\affiliation{\Demokritos}
\author{    E.~Georgiopoulou  }\affiliation{\Patras}
\author{    I.~Giomataris  }\affiliation{\Saclay}
\author{    S.~Gninenko  }\affiliation{\INR}
\author{    H.~G\' omez  }\altaffiliation[Present addr.: ]{\LAL}\affiliation{\Zaragoza}
\author{    M.~G\' omez~Marzoa  }\altaffiliation[Also at.: ]{\EPFL}\affiliation{\CERN}
\author{    E.~Gruber  }\affiliation{\Freiburg}
\author{    T.~Guth\"orl  }\affiliation{\Freiburg}
\author{    R.~Hartmann  }\altaffiliation[Present addr.: ]{\PNSensor}\affiliation{\MPI}
\author{    S.~Hauf  }\altaffiliation[Present addr.: ]{\XFEL}\affiliation{\Darmstadt}
\author{    F.~Haug  }\affiliation{\CERN}
\author{    M.~D.~Hasinoff  }\affiliation{\Vancouver}
\author{    D.~H.~H.~Hoffmann  }\affiliation{\Darmstadt}
\author{    F.~J.~Iguaz  }\affiliation{\Zaragoza}\affiliation{\Saclay}
\author{    I.~G.~Irastorza  }\affiliation{\Zaragoza}
\author{    J.~Jacoby  }\affiliation{\Frankfurt}
\author{    K.~Jakov\v ci\' c  }\affiliation{\Zagreb}
\author{    M.~Karuza  }\altaffiliation[Present addr.: ]{\Rijeka}\affiliation{\Trieste}
\author{    K.~K\"onigsmann  }\affiliation{\Freiburg}
\author{    R.~Kotthaus  }\affiliation{\MPP}
\author{    M.~Kr\v{c}mar  }\affiliation{\Zagreb}
\author{    M.~Kuster  }\altaffiliation[Present addr.: ]{\XFEL}\affiliation{\MPE}\affiliation{\Darmstadt}
\author{    B.~Laki\'{c}  }\affiliation{\Zagreb}
\author{    P.~M.~Lang  }\affiliation{\Darmstadt}
\author{    J.~M.~Laurent  }\affiliation{\CERN}
\author{    A.~Liolios  }\affiliation{\Thessaloniki}
\author{    A.~Ljubi\v{c}i\'{c}  }\affiliation{\Zagreb}
\author{    G.~Luz\'on  }\affiliation{\Zaragoza}
\author{    S.~Neff  }\affiliation{\Darmstadt}
\author{    T.~Niinikoski  }\altaffiliation[Present addr.: ]{\ECU}\affiliation{\CERN}
\author{    A.~Nordt  }\altaffiliation[Present addr.: ]{\ESS}\affiliation{\MPE}\affiliation{\Darmstadt}
\author{    T.~Papaevangelou  }\affiliation{\Saclay}
\author{    M.~J.~Pivovaroff  }\affiliation{\LLNL}
\author{    G.~Raffelt  }\affiliation{\MPP}
\author{    H.~Riege  }\affiliation{\Darmstadt}
\author{    A.~Rodr\' iguez  }\affiliation{\Zaragoza}
\author{    M.~Rosu  }\affiliation{\Darmstadt}
\author{    J.~Ruz  }\affiliation{\CERN}\affiliation{\LLNL}
\author{    I.~Savvidis  }\affiliation{\Thessaloniki}
\author{    I.~Shilon  }\altaffiliation[Also at.: ]{\BenGurion}\affiliation{\Zaragoza}\affiliation{\CERN}
\author{    P.~S.~Silva  }\affiliation{\CERN}
\author{    S.~K.~Solanki  }\altaffiliation[Sec. Affiliation: ]{\Korea}\affiliation{\MPIS}
\author{    L.~Stewart  }\affiliation{\CERN}
\author{    A.~Tom\' as  }\affiliation{\Zaragoza}
\author{    M.~Tsagri  }\affiliation{\Patras}\affiliation{\CERN}
\author{    K.~van~Bibber  }\altaffiliation[Present addr.: ]{\Berkeley}\affiliation{\LLNL}
\author{    T.~Vafeiadis  }\affiliation{\CERN}\affiliation{\Thessaloniki}\affiliation{\Patras}
\author{    J.~Villar  }\affiliation{\Zaragoza}
\author{    J.~K.~Vogel  }\affiliation{\Freiburg}\affiliation{\LLNL}
\author{    S.~C.~Yildiz  }\altaffiliation[Present addr.: ]{\Bogazici}\affiliation{\Dogus}
\author{    K.~Zioutas  }\affiliation{\CERN}\affiliation{\Patras}

\collaboration{CAST Collaboration} \noaffiliation

\begin{abstract}
The CERN Axion Solar Telescope (CAST) has finished its search for
solar axions with $^3$He buffer gas, covering the search range
$0.64~{\rm eV}\alt m_a\alt 1.17~{\rm eV}$. This closes the gap to
the cosmological hot dark matter limit and actually overlaps with
it. From the absence of excess X-rays when the magnet was pointing
to the Sun we set a typical upper limit on the axion-photon coupling
of $g_{a\gamma}\alt\hbox{3.3}\times 10^{-10}~{\rm GeV}^{-1}$ at 95\%
CL, with the exact value depending on the pressure setting. Future direct solar axion
searches will focus on increasing the sensitivity to smaller values
of $g_{a\gamma}$, for example by the currently discussed next
generation helioscope IAXO.
\end{abstract}


\maketitle


{\em Introduction.}---The most promising method to search for axions
and axion-like particles (ALPs) \cite{Jaeckel:2010ni,
Ringwald:2012hr, Hewett:2012ns, Brun:2013eia}, low-mass bosons with a two-photon
interaction vertex, is their conversion to photons in macroscopic
magnetic fields~\cite{Sikivie:1983ip, Raffelt:1987im,
Asztalos:2006kz}. This approach includes the search for solar axions
by the helioscope technique \cite{Lazarus:1992ry, Moriyama:1998kd,
Inoue:2002qy, Inoue:2008zp, Zioutas:2004hi, Andriamonje:2007ew,
Arik:2008mq, Arik:2011rx}, photon regeneration experiments (``shining
light through a wall'') \cite{VanBibber:1987rq, Redondo:2010dp,
Bahre:2013ywa}, axion-photon conversion in astrophysical $B$ fields
\cite{Payez:2012vf, DeAngelis:2011id, Horns:2012kw, Meyer:2013pny},
and the search for galactic axion dark matter \cite{Asztalos:2009yp,
Hoskins:2011iv, Asztalos:2011bm, Baker:2011na, Horns:2012jf}.

\begin{figure}
\includegraphics[width=1.\columnwidth]{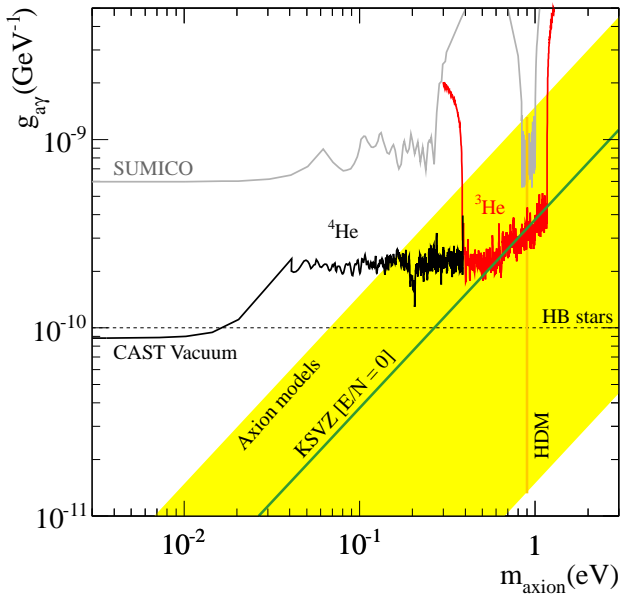}
\caption{Exclusion regions in the $m_a$--$g_{a\gamma}$--plane achieved
by CAST in the vacuum \cite{Zioutas:2004hi,Andriamonje:2007ew},
$^4$He \cite{Arik:2008mq}, and the first part of the $^3$He phase
\cite{Arik:2011rx} and our new results (all in red).
We also show constraints from Sumico
\cite{Moriyama:1998kd,Inoue:2002qy,Inoue:2008zp}, horizontal branch
(HB) stars~\cite{Raffelt:2006cw} (a somewhat more restrictive limit stems
from blue-loop suppression in massive stars~\cite{Friedland:2012hj}), and
the hot dark matter (HDM) bound~\cite{Hannestad:2010yi}. The yellow
band represents typical theoretical models with
$\left|E/N-1.95\right|=0.07$--7. The green solid line corresponds to
$E/N=0$ (KSVZ model).}\label{fig:limits}
\end{figure}

One limiting factor in any of these efforts is the momentum
difference between freely propagating photons and axions caused by
the axion mass $m_a$. It limits the magnetic field volume over which
the conversion is coherent. In solar axion searches one can extend
the search to larger $m_a$ values by providing the photons with a
refractive mass~\cite{vanBibber:1988ge}. The conversion
pipe is filled with a low-$Z$ buffer gas; the search mass is chosen by
adjusting the gas pressure. In this way, the CERN Axion Solar
Telescope (CAST), the largest axion helioscope to date, has
successively pushed its search range to higher $m_a$ values (see
Fig.~\ref{fig:limits} for a summary of results). We here report on
the final search range based on $^3$He buffer gas.

Within the ALP family of hypothetical bosons, the original axion is
the best-motivated case because it emerges from the compelling
Peccei-Quinn mechanism to explain the absence of CP-violating
effects in QCD. In the two-dimensional $g_{a\gamma}$-$m_a$ ALP parameter space, the QCD axion must lie somewhere on a line $g_{a\gamma}\propto m_a$. The close relationship between axions and neutral pions implies that this line is anchored to the point describing the $\pi^0$ mass and the pion-photon coupling constant. After
allowing for model-dependent numerical factors, the axion may be
found anywhere in the yellow band indicated in
Fig.~\ref{fig:limits}. The CAST vacuum result ($g_{a\gamma}<0.88\times10^{-10}~{\rm GeV}^{-1}$ at 95\% CL for
$m_a\alt 0.02$~eV \cite{Andriamonje:2007ew}) remains a milestone in
the ALP landscape. However, a major objective of CAST has been to find or exclude QCD axions and thus to push as far as possible to higher $m_a$ values. Our first $^3{\rm He}$ limits
\cite{Arik:2011rx} have for the first time crossed the axion line
appropriate for the Kim, Shifman, Vainshtein, Zakharov (KSVZ) model (Fig.~\ref{fig:limits}) \cite{Kim:1979if,Shifman:1979if}.

QCD axions with parameters in this range thermalize in the early
universe after the QCD phase transition by interactions with pions
\cite{Chang:1993gm} and would thus exist with a present-day number
density of around $50~{\rm cm}^{-3}$, comparable to 0.5 neutrino
species, and are therefore susceptible to hot dark matter bounds
\cite{Hannestad:2005df, Melchiorri:2007cd, Hannestad:2010yi}.
Assuming neutrino masses to be negligible, the latest axion hot dark
matter bound is $m_a\alt0.9$~eV, leaving a small gap to our earlier
$^3$He search range which we now close.

The recent Planck measurements of the cosmic microwave background
(CMB) significantly improve our knowledge of many cosmological parameters.  In contrast to earlier CMB results, Planck alone now constrains the axion mass and provides a limit $m_a<1.01$~eV (95\% CL)~\cite{Archidiacono:2013cha}. The inclusion of other data sets, notably the matter power spectrum and the HST measurement of the Hubble parameter, have only a small impact, providing limits between 0.67 and 0.86~eV, depending on the combination of data sets~\cite{Archidiacono:2013cha}.  In other words, concerning a possible axion hot dark matter contribution to the universe, the situation after Planck is almost the same as before.

{\em System description and data-taking strategy.}---CAST uses a straight 10m LHC test dipole magnet (B$\sim$ 9.0~T), mounted on a movable platform to follow the Sun for about 1.5~h both at sunrise and sunset. The two bores extend beyond the cold mass (length 10.25~m) for 16~cm on each side forming 4 ‘link’ regions which are closed by x-ray cold windows. The volume of the two cold bores is 30~L and the total volume of the link regions is 1.5~L. The magnetic field length of 9.26~m is centrally located within the cold mass. One of the apertures of the magnet is covered by a CCD/Telescope system
\cite{Kuster:2007ue} and the other three by three Micromegas detectors of the microbulk type \hbox{\cite{Abbon:2007ug,Andriamonje:2010zz, Galan:2010zz, Aune:2009zzc}}. 
The axion-photon conversion probability when the conversion volume is filled with a buffer gas ($^{3}$He in our case) is \cite{Arik:2008mq}
\begin{equation}
P_{a \rightarrow \gamma}    =  \left( \! \frac{Bg_{a \gamma}}{2} \! \right)^{2}
       \,\, \frac{1 \! + \! e^{-\Gamma L} \! - \! 2 e^{-\Gamma L/2}
 \cos(qL)}{q^{2} \! + \! \Gamma^{2} \! /4}
        \label{prob}
\end{equation}
\noindent where the axion-photon momentum transfer provided by the magnetic field is $q=|m_a^2-m_\gamma^2|/2E$ and $\Gamma$ is the inverse photon absorption length in the buffer gas. The value of $\Gamma$ varies with the pressure and the energy, for example for a relatively high pressure of 70~mbar of $^3$He, for the mean energy of the expected flux of 4.3~keV, $\Gamma=0.156~{\rm m}^{-1}$. The maximum conversion probability is reached for $m_a \simeq m_\gamma$ where $m_{\gamma}$ is the photon refraction mass which depends on the buffer gas density. For $m_a\not=m_\gamma$, the probability rapidly decreases due to the axion-photon momentum mismatch.

Throughout CAST Phase II, the data taking strategy was to increase the density in the cold bore circuit in small steps chosen to partially overlap the intrinsic mass acceptance ($\sim$1~meV FWHM) of the previous setting and so scan smoothly over the whole available mass range. The original step size and exposure time have been modified on a number of occasions in order to complete the physics program more efficiently without compromising continuity, but at the expense of reduced sensitivity at higher masses.

The central gas density inside the cold bore, with the magnet horizontal, is calculated from the cold bore pressure ($P_{\rm cb}$) measured at one end, the magnet temperature $T_{\rm mag}$ and the equation of state (EoS) of $^3$He gas \cite{lemmon}. During solar tracking, $P_{\rm cb}$ changes continuously, as expected, due to the changing hydrostatic pressure of the $^3$He gas column and due to a tilt-induced, slow characteristic temperature transient in the magnet (10--15~{\rm mK}) from the cryogenic circuit. For example, at $P_{\rm cb}$ 84~mbar, a vertical movement of the magnet of 6~degrees causes a shift in $P_{\rm cb}$ of +1.06~mbar. Hydrostatic and magnet temperature effects account for +0.65~mbar. The remaining contribution of 0.4~mbar we ascribe to changing fluid dynamics in the buffer gas at the extremities of the cold bore.

The fluid dynamics effect is driven by the presence of short relatively warm link regions; the $^3$He temperature and density are not uniform throughout the whole system as regions with lighter vapor are present at the extremities, where buoyancy-driven flows occur. The magnet tilting affects such phenomena, giving rise to a redistribution of the $^3$He mass and a consequent pressure change. 
To take the pressure and temperature variation into account, our analysis procedure continuously calculates the density during solar tracking. First, the pressure in the center of the magnet is calculated from the $P_{\rm cb}$ and the hydrostatic pressure difference. Then the density in the center is calculated from this central pressure and the temperature measurement (via the $^3$He EoS). In this way the fluid dynamics effects on the measured pressure directly change the central density value.

Although the $P_{\rm cb}$ measurement allows us to calculate the central density at any moment, the actual density profile (which is needed to calculate the coherence length) and its variation on tilting cannot be measured directly and must be determined by Computational Fluid Dynamics (CFD) simulations. The CFD simulations take into account all requisite physical phenomena, such as gravity, natural convection and turbulence together with the geometry of the cold bores, link volumes and the cold windows and the buffer gas EoS. The boundary conditions are defined by $P_{\rm cb}$, $T_{\rm mag}$  and several temperatures measured around the link volumes and cold window flanges.

An extensive and on-going program of CFD simulations has been undertaken and CAST has made detailed studies with a number of different models to find the best description of the measured behavior. The tilted and horizontal cases were treated separately. Various turbulence models were used for the horizontal case and a model forcing laminar flow was favored, while a composite model was devised for the tilted case as the most intuitive natural description of the system. This model consists of a turbulent solution in the lower half of the cold bore smoothly joined to a laminar solution in the upper half. The predicted pressure variations between tilted simulations at different vertical angles are in satisfactory agreement with those observed experimentally (e.g. within 0.06~mbar for 84~mbar.) 

For the analysis presented in this paper, the density profiles derived from  turbulent CFD simulations made with the magnet horizontal and over the full range of Phase II density settings were subjected to a simple and conservative coherence criterion  ($\Delta \rho<0.001~{\rm kg m}^{−3}$). The resulting dependence of the effective coherence length $L_{\rm eff}$ with density was parametrised and applied to all data independent of photon energy and tilt angle. $L_{\rm eff}$ decreases from about $\sim9~{\rm m}$ to $\sim6~{\rm m}$  in the range $m_a$=$0.4~{\rm eV}$ to $m_a$=$1.15~{\rm eV}$, compared with the magnetic length of $9.26~{\rm m}$. To estimate the systematic error of such an approach, an analysis was done using a coherence length $L_{\rm eff}$= 5.0~m for all angles and masses. This extreme case is only found in laminar horizontal simulations at the highest pressures. The final effect on the limit on the $g_{a\gamma}$ from applying this simple criterion is well below 10\%.


{\em Data analysis and results.}---The results presented in this paper correspond to 1100~hour$\times$detector taken by the three Micromegas detectors from 2009 to 2011 with $^3$He in the system in axion-sensitive conditions (i.e. with the magnet tracking the Sun). Background levels are determined from a larger body of data taken during non tracking time. The data acquired by the CCD/Telescope of this period is under analysis and will be presented in a later publication. The present data correspond to about 418 effective axion mass steps that, together with the first 252 $^3$He steps already released in a previous paper \cite{Arik:2011rx}, continuously cover an axion mass range between 0.39~eV and 1.17~eV. Due to the density excursions experienced during a single tracking, data from each actual density step contribute to the neighbouring mass steps, especially for the larger densities used. The effective average exposure time per mass step is  approximately 0.75~h per detector for masses from 0.64~eV to 1~eV, while it was reduced to $\sim$0.4~h per detector for masses above 1~eV. 

The data analysis is performed in a manner similar to our previous
results~\cite{Zioutas:2004hi,Andriamonje:2007ew,Arik:2008mq,Arik:2011rx}.
We use an unbinned likelihood function that can be expressed as
\begin{equation}\label{unbinned}
\log \mathcal{L} \propto -R_T + \sum_i^N \log R(t_i, E_i, d_i)\,.
\end{equation}
Here $R_T$ is the integrated expected number of counts over all exposure time, energy and detectors. The sum runs over each of the $N$ detected counts for the event rate $R(t_i,
E_i, d_i)$ expected at the time $t_i$, energy $E_i$
and detector $d_i$ of the event $i$ 
\begin{equation}\label{R}
R(t, E, d) = B_d + S(t,E,d)\;,
\end{equation}
where $B_d$ is the background rate of detector $d$. $S(t,E,d)$ is the
expected rate from axions in detector $d$ which depends on the axion
properties $g_{a\gamma}$ and $m_a$
\begin{equation}\label{S}
S(t,E,d) = \frac{d\Phi_a}{dE} P_{a\rightarrow \gamma} \epsilon_d\;,
\end{equation}
where  $P_{a\rightarrow \gamma}$ is the axion photon conversion
probability in the CAST magnet given by Eq.~(\ref{prob}) and
$\epsilon_d$ the detector effective area. Finally, the solar axion
spectrum based on the Primakoff process is the same that was used in previous papers of this series~\cite{Andriamonje:2007ew} 

\begin{equation}
 \frac{d\Phi_a}{dE}  =  6.02 \times 10^{10} \,  g_{10}^{2} \,
  \frac{E^{2.481}}{e^{E/1.205}} \, \, \, \, \, \,  \mathrm{cm}^{-2} \, \mathrm{s}^{-1} \, \mathrm{keV}^{-1}
\label{spectrum}
\end{equation}
with $g_{10}~=~g_{a \gamma}/(10^{-10} \, \mathrm{GeV}^{-1})$ and
energies in keV. This result applies to axions with masses much smaller than the solar interior temperature, i.e., for sub-keV masses.

As explained in \cite{Arik:2008mq}, the $m_a$ dependence of the above
expression is encoded in the probability $P_{a\rightarrow \gamma}$,
which is coherently enhanced for values of $m_a$ matching the
refractive photon mass $m_\gamma$ induced by the buffer gas density,
while it is negligible for values away from $m_\gamma$. Therefore,
only the counts observed with the gas density matching a given axion
mass $m_a$ will contribute to $\log \mathcal{L}$ (and the exclusion plot) for
that mass $m_a$. We stress that the value of $m_\gamma$ to be
introduced is time-dependent even within a single density step, due
to the pressure excursions explained above.

\begin{figure}
\includegraphics[width=1.\columnwidth]{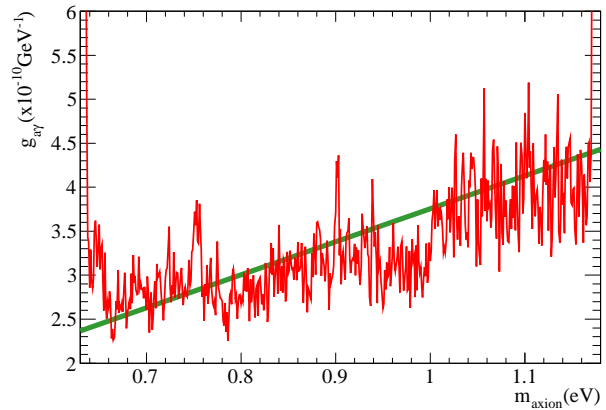}
\caption{Expanded view of the limit achieved in the CAST $^3$He phase for the axion mass range between 0.64 eV and 1.17 eV, which corresponds to a pressure scan in $^3$He from 36~mbar to 105~mbar approximately. The actual limit contour has a high-frequency structure that is
a result of statistical fluctuations that occur when a limit is computed for a specific mass using only a few hours of data. The green solid line corresponds to $E/N=0$ (KSVZ model).}\label{fig:zoom}
\end{figure}

Maximization of $\mathcal{L}$ (for a fixed value of $m_a$) leads to a best-fit
value of $g_{\mathrm{min}}^4$. The obtained value is compatible with
the absence of a signal in the entire mass range, and therefore an
upper limit on $g_{95}^4$ is obtained by integration of the Bayesian
probability from zero up to 95\% of its area in $g^4$. This is
computed for many values of the axion mass $m_a$ in order to
configure the full exclusion plot shown in Fig.~\ref{fig:limits}. A
close up of the same exclusion plot is shown in Fig.~\ref{fig:zoom},
focused specifically in the axion mass range which has been explored
in the data presented here. 

As can be seen in Fig.~\ref{fig:limits}, CAST extends its previous
range towards higher axion masses, excluding the interval
0.64--1.17~{\rm eV} down to an average value of the axion-photon
coupling of $3.3\times 10^{-10}~{\rm GeV}^{-1}$. The actual limit
contour has a high-frequency structure that is a result of statistical
fluctuations that occur when a limit is computed for a specific mass
using only a few hours of data. The apparent slope upwards in the exclusion line for higher axion masses is due to the reduction of the exposure time per density step, for  $m_a>1$~eV, as well as to the continuous decrease of $L_{\rm eff}$ and the increase of $\Gamma$ for higher gas densities. Eventually, with the addition of the
data from the CCD/Telescope system, these numbers will likely
improve.

{\em Conclusions.}---CAST has finished its phase of using $^3$He
buffer gas, continuing the search to its limiting pressure setting
corresponding to a search mass of $m_a=1.17$~eV. In this way, the
search range now generously overlaps with the current cosmic hot dark
matter bound of $m_a\alt 0.9$~eV and there would be little benefit in
pushing to yet larger masses with the buffer-gas technique. CAST has
not found axions and the next challenge is to move down in the
$m_a$--$g_{a\gamma}$ plot to reach the ``axion band'' of theoretical
models in a broader range of masses. Such a goal cannot be achieved
with the existing CAST apparatus and will require significant
improvements of detector and magnet
properties, such as the proposed International AXion Observatory (IAXO)~\cite{Irastorza:2011gs, Shilon:2012} or a completely new
approach.

{\em Acknowledgments.}---We thank CERN for hosting the experiment and
for the technical support to operate the magnet and cryogenics. We
thank the CERN CFD team for their essential contribution to the CFD
work. We acknowledge support from NSERC (Canada), MSES (Croatia)
under the grant number 098-0982887-2872, CEA (France), BMBF (Germany)
under the grant numbers 05 CC2EEA/9 and 05 CC1RD1/0 and DFG (Germany)
under grant numbers HO 1400/7-1 and EXC-153, GSRT (Greece),
NSRF: Heracleitus II, RFFR (Russia), the Spanish Ministry of Economy and Competitiveness
(MINECO) under grants FPA2008-03456 and FPA2011-24058. This work was partially funded by the European Regional Development Fund (ERDF/FEDER), the European Research Council (ERC) under grant ERC-2009-StG-240054 (T-REX), Turkish Atomic
Energy Authority (TAEK), NSF (USA) under Award number 0239812, US
Department of Energy, NASA under the grant number NAG5-10842. Part of
this work was performed under the auspices of the US Department of
Energy by Lawrence Livermore National Laboratory under Contract
DE-AC52-07NA27344. We acknowledge the helpful discussions within the
network on direct dark matter detection of the ILIAS integrating
activity (Contract number: RII3-CT-2003-506222).


\end{document}